\documentclass[traditabstract]{aa}  
\pdfoutput=1

\usepackage[usenames,dvipsnames,svgnames,table]{xcolor}
\usepackage{graphicx}
\usepackage[varg]{txfonts}
\usepackage{epstopdf}
\usepackage{xspace} 
\usepackage{natbib}
\bibpunct{(}{)}{;}{a}{}{,}

\newcommand{\spitzer}{\textit{Spitzer}\xspace}
\newcommand{\herschel}{\textit{Herschel}\xspace}

\newcommand{\degree}{\mbox{$^{\circ}$}\xspace}

\newcommand{\micron}{\ensuremath{\mu\mathrm{m}}}
\newcommand{\percc}{\ensuremath{\textrm{cm}^{-3}}}
\newcommand{\kms}{\textrm{km~s}\ensuremath{^{-1}}}
\newcommand{\msunyr}{M$_\odot$~yr$^{-1}$}% Msun per year
\newcommand{\lsun}{\ensuremath{L_{\odot}}}			%  Lsun
\newcommand{\lbol}{\ensuremath{L_{\mathrm{bol}}}}	 %  Lbol
\newcommand{\water}{H$_2$O}
\newcommand{\macc}{\ensuremath{\dot{M}_{\mathrm {acc}}}}	
\newcommand{\lacc}{\ensuremath{L_{\mathrm{acc}}}}

\begin{document}
  \title{\herschel/PACS far-IR spectral imaging of a jet from an intermediate mass protostar in the OMC-2 region \thanks{ {\it \herschel} is an ESA space observatory with science instruments provided by European-led Principal Investigator consortia and with important participation from NASA.} }

\titlerunning{[O I] jet from OMC-2 FIR~3}
\authorrunning{B. Gonz\'alez-Garc\'ia et al.}

   \author{B. Gonz\'alez-Garc\'ia
          \inst{1} \ \inst{2} \and P. Manoj \inst{3} \and D. M. Watson \inst{4}
	  \and R. Vavrek \inst{1} \and S. T. Megeath \inst{5}  \and A. M. Stutz \inst{6}   \and M. Osorio \inst{7} \and F. Wyrowski \inst{8} \and W. Fischer \inst{9}   \and  J. J. Tobin \inst{10}  \and M. S{\'a}nchez-Portal \inst{1} \ \inst{2}  \and  A.K. Diaz Rodriguez \inst{7} \and T. L. Wilson \inst{11}   }
	 
%\fnmsep\thanks{Just to show the usage of the elements in the author field}
	  
\offprints{bgonzale@sciops.esa.int}

   \institute{European Space Astronomy Centre, ESA, P.O. Box 78, 28691 Villanueva de la Ca\~nada (Madrid), Spain, \\
   \email{bgonzale@sciops.esa.int}, \and ISDEFE, Beatriz de Bobadilla 3, 28040, Madrid, Spain \and Tata Institute of Fundamental Research, Homi Bhabha Road, Colaba, Mumbai 400005, India,  \and  Department of Physics and Astronomy, University of Rochester, Rochester, NY 14627, USA, \and Department of Physics and Astronomy, University of Toledo, 2801 West Bancroft Street, OH 43606, USA,\and  Max-Planck-Institute for Astronomy, K\"onigstuhl 17, 69117 Heidelberg, Germany   \and Instituto de Astrof\'isica de Andaluc\'ia, CSIC, Camino Bajo de H\'uetor 50, E-18008 Granada, Spain \and Max-Planck-Institut f\"ur Radioastronomie, Auf dem H\"ugel 69, 53121, Bonn, Germany  \and NASA Goddard Space Flight Center,  Greenbelt, MD, USA  \and    Leiden Observatory, Leiden University, P.O. Box 9513, 2300-RA Leiden, The Netherlands  \and Naval Research Laboratory, Washington, DC, USA}
\date{\it Accepted for publication in A\&A}

  \abstract{We present the first detection of a jet in the far-IR [O~I] lines from an intermediate mass protostar. We have carried out a \herschel/PACS spectral mapping study in the [O~I] lines of OMC-2 FIR~3 and FIR~4, two of the most luminous protostars in Orion outside of the Orion Nebula. The spatial morphology of the fine structure line emission reveals the presence of an extended photodissociation region (PDR) and a narrow, but intense jet connecting the two protostars. The jet seen in [O~I] emission is spatially aligned with the \spitzer/IRAC 4.5~\micron~jet and the CO (6-5) molecular outflow centered on FIR~3. 
% The [O I] 63~\micron~to~145~\micron~line ratios indicate a density gradient along the jet with density decreasing progressively as one moves away from FIR~3 on either side. 
  The mass loss rate derived from the total [O~I] 63 \micron~line luminosity of the jet is 7.7~$\times$10$^{-6}$~\msunyr, more than an order of magnitude higher than that measured for typical low mass class 0 protostars.  The implied accretion luminosity is significantly higher than the observed bolometric luminosity of FIR~4,  indicating that the [O~I] jet is unlikely to be associated with FIR~4. We argue that the peak line emission seen toward FIR~4 originates in the terminal shock produced by the jet driven by FIR~3.  
  %From the observed upper limit of the [O~III]~88.4~\micron~line flux we place a limit on the mass of the ionised gas around FIR~4, thus ruling out the presence of an H~II region previously suggested to coincide with FIR~4. 
  The higher mass-loss rate that we find for FIR~3 is consistent with the idea that intermediate mass protostars drive more powerful jets than their low-mass counterparts.  Our results also call into question the nature of FIR~4.}
  

\maketitle

\section{Introduction}

The far-infrared lines of [O~I] are prominent tracers of  jets from young stars and protostars. They provide a straightforward measure of the mass flow through dissociative J-shocks driven by the outflowing jets  \citep{werner84, cohen88, nisini96, hollen85, hm89}. In regions of high extinction, as those found around young protostars, [O~I] lines are a means to trace and characterize jets from outflows, unaffected by extinction or confusion with the ambient molecular cloud. While [O~I] jets from optically revealed pre-main sequence stars and  low-mafss protostars have been detected with \herschel/PACS \citep[e.g.,][]{podio12, nisini15}, no such detections have been reported for embedded intermediate mass protostars.

\begin{table*}
{\small
\centering
%Table type Manoj paper !!!
\caption{Log of PACS spectroscopic observations \label{log} } % title of Tab
% is used to refer this table in the text 
%\begin{sidewaystable}
%\begin{center}
\begin{tabular}{l l r c r c c c c c } % centered columns (4 colum
\hline 
HOPS ID & ObsID & OD & Date &Total time & RA & DEC & Observing mode & Spectral range \\ % table heading 
& & & & (s) & ($\deg$)& ($\deg$)& & \\ \hline % inserting body of the table 
FIR~3 (OFF) & 1342227763 & 838 & 2011 Aug 30 & 989 & 5 32 18.750& $-$6 24 27.50 & pointed & 57-71 $\&$170-213 \\
FIR~3 (ON) & 1342227764 & 838 & 2011 Aug 30 & 8933 & 5 35 27.670 & $-$5 09 35.10 & $3 \times 3$ raster & 57-71 $\&$170-213 \\
FIR~3 (OFF) & 1342227765 & 838 & 2011 Aug 30 & 989 & 5 32 18.750& $-$6 24 27.50 & pointed & 57-71 $\&$170-213 \\
FIR~3 (ON) & 1342227766 & 838 & 2011 Aug 30 & 8933 & 5 35 27.670 & $-$5 09 35.10 & $3 \times 3$ raster & 57-71 $\&$170-213 \\
FIR~3 (OFF)& 1342227768 & 838 & 2011 Aug 30 & 1520 & 5 32 18.750& $-$6 24 27.50 & pointed &50-98 $\&$102-196 \\
FIR~3 (ON)& 1342227769 & 838 & 2011 Aug 30 & 6083 & 5 35 27.670& $-$5 09 35.10 & $2 \times 2$ raster &50-98 $\&$102-196 \\
FIR~3 (OFF) & 1342227770 & 838 & 2011 Aug 30 & 1520 & 5 32 18.750& $-$6 24 27.50 & pointed & 50-98 $\&$102-196 \\
FIR~3 (ON)& 1342227771 & 838 & 2011 Aug 30 & 6083 & 5 35 27.670 & $-$5 09 35.10 & $2 \times 2$ raster &50-98 $\&$102-196 \\
\\
FIR~4 (ON)& 1342239690 & 1020 & 2012 Feb 27 & 1930 & 5 35 27.070 & $-$5 10 00.40 & pointed &50-71 $\&$102-142 $\&$ 170-213 \\
FIR~4 (OFF) & 1342239691 & 1020 & 2012 Feb 27 & 1930& 5 32 18.750& $-$6 24 27.50 & pointed &50-71 $\&$102-142 $\&$ 170-213 \\
FIR~4 (OFF) & 1342239692 & 1020 & 2012 Feb 27 & 3079 & 5 32 18.750& $-$6 24 27.50 & pointed & 71-98 $\&$141-196 \\
FIR~4 (ON)& 1342239693 & 1020 & 2012 Feb 27 & 3079 & 5 35 27.070 & $-$5 10 00.40 & pointed & 71-98 $\&$141-196 \\\hline
\hline
\end{tabular}
%\end{sidewaystable}
%\end{center}
}
\end{table*}

\begin{figure*}
\centering
\begin{tabular}{ccc}
\includegraphics[scale=0.3]{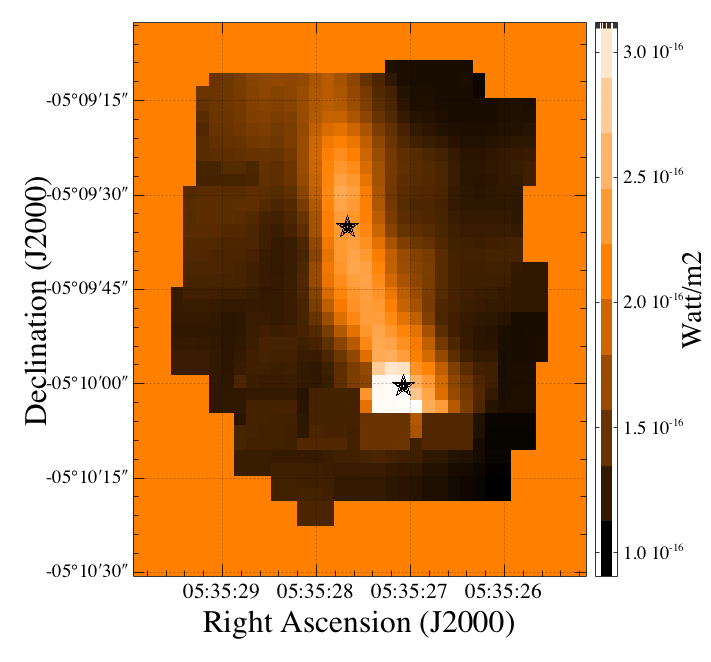} & &
\includegraphics[scale=0.3]{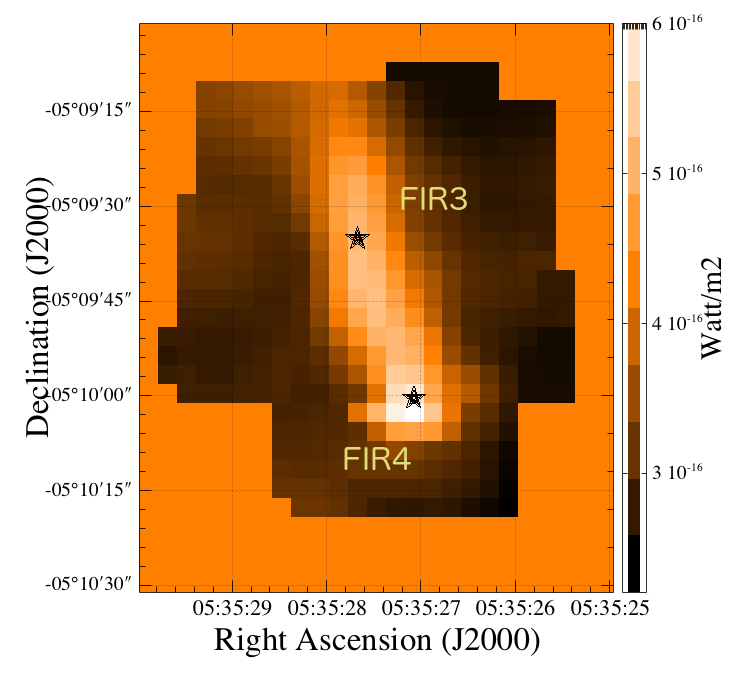} \\
\end{tabular}
\caption{ {\it (Left)}:  [O~I] 63~\micron~line map generated with Drizzle and specInterpolate for 2\arcsec.  {\it (Right)}:  [O~I] 63~\micron~line map generated with specProject and specInterpolate for 3\arcsec.  The left image provides a sharper spatial reconstruction north to $-$05$^{\circ}$~10\arcmin~of the FIR~3 jet component by taking advantage of drizzling but the right image has been used for this study allowing in order to achieve a homogeneous and consistent mapping of the entire area.
 \label{figa1}}
\end{figure*}

\begin{figure*}
\begin{tabular}{ll}
\centering
\includegraphics[scale=0.33]{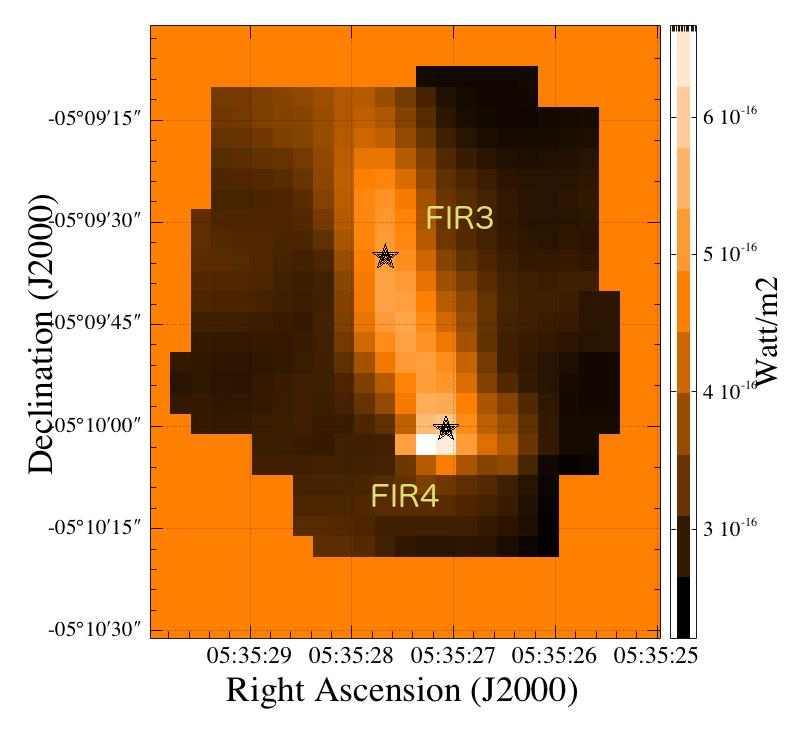} & 
\includegraphics[scale=0.33]{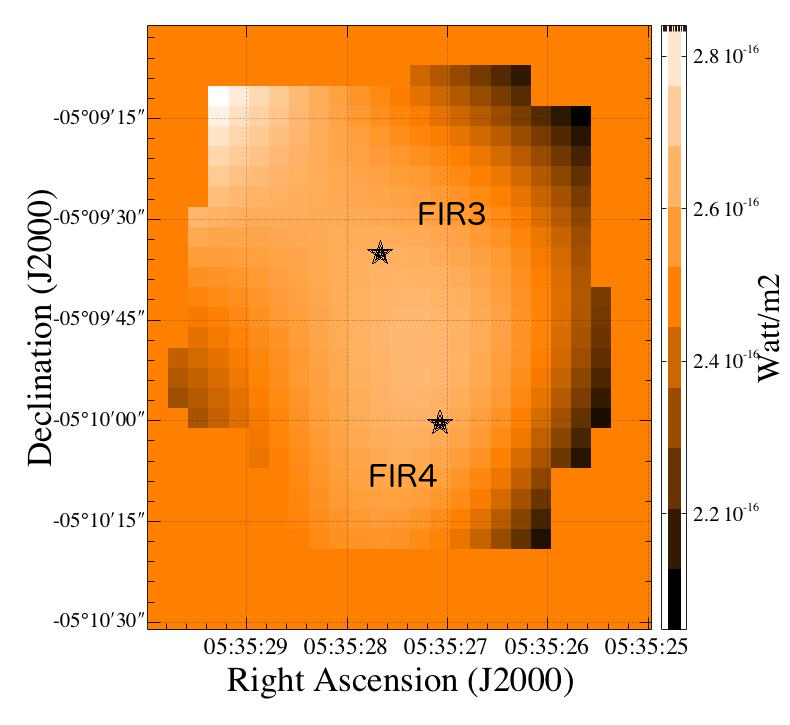} \\
\includegraphics[scale=0.5]{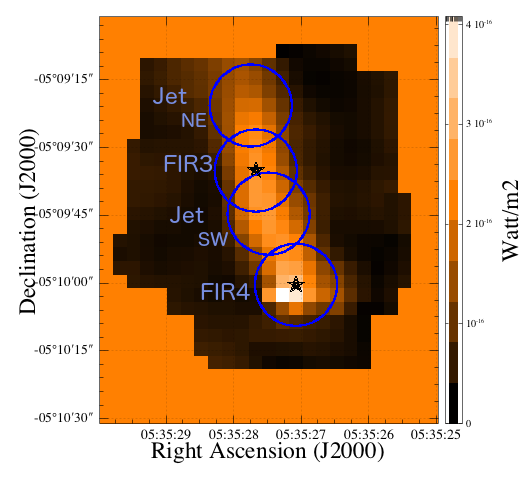} &  
\end{tabular}
\caption{ {\it (Top left)}:  [O~I] 63~\micron~line map of the OMC-2 region.  {\it (Top right)}:  Extended emission component extracted by fitting a third order polynomial to the line map. {\it (Bottom)}: Narrow ridge like emission component obtained after subtracting the extended emission background component. Positions, sizes, and names of the line flux aperture measurements listed in Table~2 are indicated. 
%The value of the flux was of $\sim$ of 5.5$\times$ 10$^{-16}$ Watt m$^{-2}$.
 \label{fig1}}
\end{figure*}

\begin{figure*}
   \centering
\begin{tabular}{cc}
\includegraphics[scale=0.34]{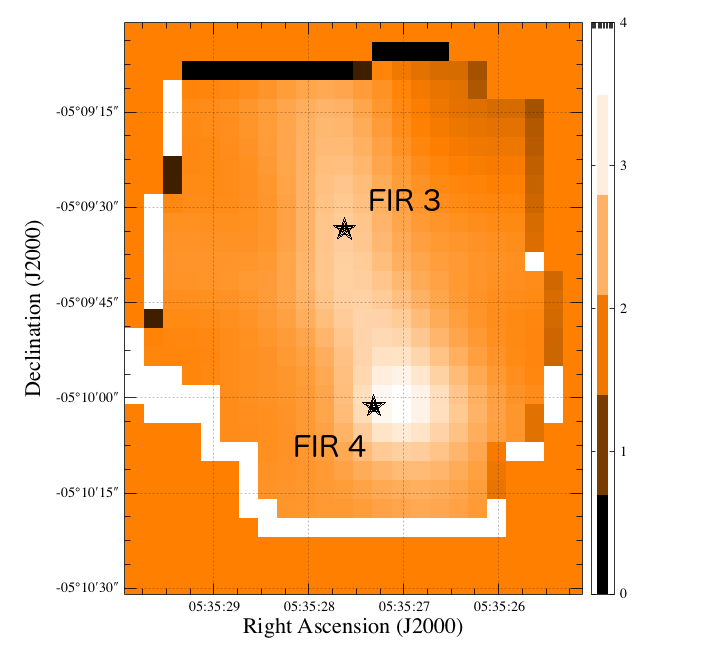} &
\includegraphics[scale=0.3]{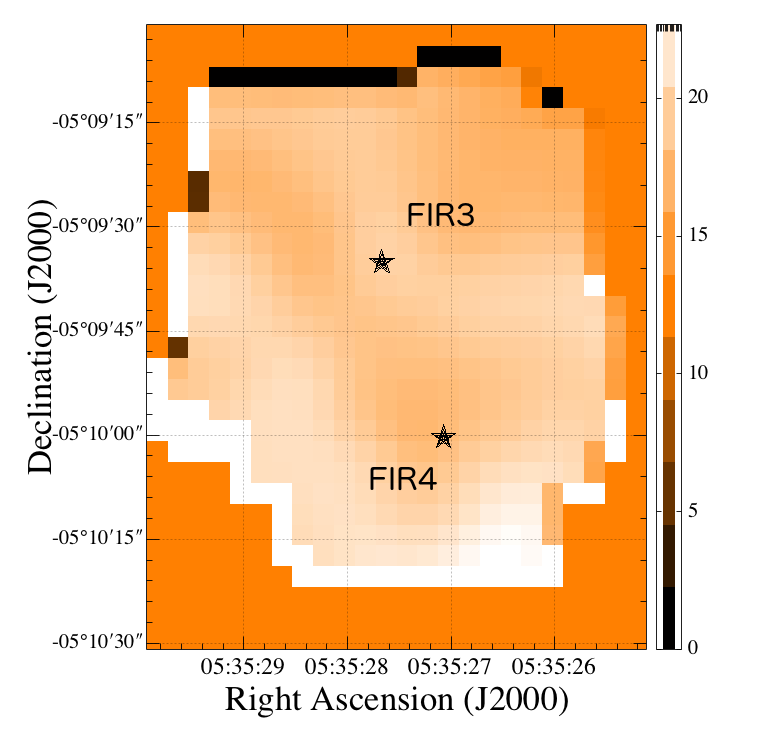} \\
\end{tabular}
\caption{ {\it (Left):} [O~I] 63~\micron~/~[C II] 158~\micron~map, {\it (Right): }  [O~I] 63~\micron~ /  [O~I] 145~\micron~map.  All line ratios are calculated at matched angular resolution. 
\label{fig_oc} }
\end{figure*}

The OMC-2 region in Orion A is one of the richest targets for  [O~I] studies of outflows from intermediate mass protostars.  It contains a group of eight protostars with luminosities of a 20 to 360~\lsun~\citep{adams12,furlan16}, much higher than the typical luminosity of 1~\lsun~for the peak of the protostellar luminosity function of the Orion protostars \citep{kryukova12,  furlan16}.  OMC-2, part of a 2 pc long, dense filament which extends between the Orion Nebula  Cluster (ONC) to the south and NGC 1977 to the north, hosts a higher than average fraction of young Class 0 protostars \citep{chini97, peterson08, adams12, megeath12,  stutz15}. 
%Due to the high luminosities of its protostars, the lack of massive stars that would both influence the gas and dominate the IR flux, and the relative proximity of Orion \citep[420~pc;][menten], OMC-2 has been
%extensively observed in the infrared (IR) and submillimeter (sub-mm) with the goal of studying intermediate-mass star formation.
%Due to the relatively high luminosities of its protostars, the lack of massive stars that both strongly influence the gas and dominate the IR flux, their youth, and their relative proximity at 420~pc, this region has been extensively observed at the IR and sub-mm wavelengths with the goal of studying intermediate mass star formation. 
%Of particular interest are the OMC-2 protostars FIR~3 (HOPS~370) and FIR~4 (HOPS~108), which are separated only by $\sim$~20\arcsec \citep[e.g.,][]{furlan14}. 

In this paper, we present data obtained as part of the Herschel  Orion Protostar Survey (HOPS) open time key program. We analyze Herschel/ PACS fine structure line observations of the OMC-2 protostars FIR 3 (HOPS 370, VLA 11) and FIR 4 (HOPS 108, VLA 12) \citep[see][]{reipurth99, adams12, furlan16}.  FIR~3 is the most luminous protostar (\lbol $= 360$~\lsun) in OMC-2 and drives a bipolar outflow with one lobe extending to FIR~4~\citep{williams03, shimajiri08}. While previous observations indicated that FIR~4 has a luminosity of 1000~\lsun~\citep{crim09}, the bolometric luminosity has now been revised to 38~\lsun~\citep[see ][for a detailed discussion]{furlan14}. Low-$J$ ($J_{up}~\le$~3) CO observations do not show a distinct outflow associated with FIR~4. FIR~3 and FIR~4 have the brightest known far-IR line emission in the cloud outside of the Orion nebula \citep{manoj13}. While FIR~3 has a far-IR CO luminosity of $\sim$~0.06~\lsun, FIR~4 has a far-IR CO luminosity of $\sim$~0.2~\lsun~and the brightest [O~I] and \water~emission observed among the HOPS sample~\citep{manoj13}.  Moreover, a clear peak in the far-IR line emission is found at the location of FIR~4 \citep{manoj13, furlan14}.  While our understanding of the nature of FIR~4 is still incomplete (see Osorio et al., in prep.), the [O~I] detailed physical nature of FIR~3 is now coming into sharper focus. 

We detect, for the first time, a jet in the far-IR [O~I] lines from an intermediate mass protostar FIR~3.  The jet line luminosity and implied mass loss rate are a factor of 10 higher than typical values for low-mass class 0 protostars.  This jet extends from FIR~3 to FIR~4 and is spatially coincident with CO (6-5) outflow observations, demonstrating a connection between FIR~3 and the bright far-IR line emission at the position of FIR~4. 
%The inferred mass loss rates and the electron densities of the [O~I] jet are entirely consistent with the intense line emission seen from FIR 4 originating in the terminal shock produced by the jet driven by FIR~3. 
We propose that the bright line emission associated with FIR~4 is produced by the FIR~3 jet.

\section{Observations \& data reduction}
%%%

\subsection{\herschel PACS}
The PACS spectroscopic observations of the OMC-2 region were carried out using the Range Spectroscopy mode to achieve a Nyquist-sampled coverage of the $57-196$~\micron~wavelength range. The region centered on FIR~3 was observed in the full PACS spatial resolution mapping mode, oversampling the beam by using a 3$\times$3 raster in the blue and 2$\times$2 raster in the red channel with sub-pixel raster line step sizes of 3$\arcsec$ and 4.5$\arcsec$, respectively. The region centered on FIR~4 was observed in the single-pointing mode. We applied the unchopped beam-modulation technique developed for crowded-field spectroscopy. In the unchopped mode, off-positions free of cloud emission within 2\degree from the target were observed to remove the telescope and sky background. 
The log of PACS observations of the OMC-2 region is given in Table~\ref{log}.
%and a detailed description of observations and the various steps involved in the data reduction is provided in Appendix~\ref{sec:observations}.

The Nyquist sampled wavelength range of $57-71$~\micron~was observed
in the third grating order (B3A), the range $71-98$~\micron~in the
second order (B2B), and the range $102-196$~\micron~in the first order
(R1). The second (B2B) and third (B3A) grating orders are observed
with the blue camera, while the first order (R1) spectra has been
taken with the red camera simultaneously with one of the blue
observations. The broad-range spectral scans were repeated 2 times on
each raster position for the mapping observations, and 4 times for the
pointed observations in order to achieve at least a 5$\sigma$ detection
for the diagnostic lines across the sampled area and to improve
spectral sampling provided by the sub-pixel offsets between grating
scans.

The HOPS 370 (FIR~3) mapping observation was composed of a sequence of
on-target raster and off-target pointed observations which ensures the
continuum signal-to-noise achieved per raster position is identical to
the signal-to-noise achieved within the single off-position
observation. The off-block is subtracted from every raster position of
the corresponding mapping observation. In order to shorten the
timescale between on- and off-blocks each of the B3A/R1 and B2B/R1
observations have been split up into two identical subgroups resulting
4 on- and off observations pairs in total, on the extra cost of
$\sim$~15 min instrument and spacecraft overheads. For the 3$\times$3
blue rasters $\sim$~89\% of time was spent-on source, while for the
2$\times$2 red rasters $\sim$~75\% of the time was used for on-target
observations exhibiting an excellent observing efficiency. The HOPS
108 (FIR~4) pointed observations followed a similar design principle
except the choice of wavelength range combinations introduces a
slightly improved redundancy and deeper detection limit. The on-source
time consists $\sim$~50\% of the total observing time of pointed
observations.

FIR~3 was observed in 2011 August over a total of 10.4 hrs
(including overheads), and FIR~4 was observed on 2012 February over
2.8 hrs. Table~\ref{log} gives the observation list, observing mode
parameters, covered wavelength ranges and individual integration
times. The \herschel nominal absolute pointing accuracies for the
periods comprising the observations of FIR~3 and FIR~4 are 1\farcs3
and 0\farcs9, respectively \citep{SanchezPortal2014}. However,
in order to improve the short-term relative pointing accuracy
(pointing stability), the attitude data were reprocessed by means of a
ground reconstruction procedure making use of the data from the
gyroscopes. This procedure improves the instrumental PSF and
gives a very uniform absolute pointing accuracy of 1\farcs2 across
the periods of the mission relevant to this paper. However, the
observations of FIR~4 were carried with a solar aspect angle (SAA)
of about 102.5\degree. At such angles, some pointing shifts due to
structural thermoelastic effects could be expected \citep{SanchezPortal2014} and hence a conservative accuracy figure of [O~I]  2\farcs0 is advisable for this second target.

%\subsection{Data reduction}
The data were reduced with the \herschel interactive processing
environment (HIPE v12.1 and calibration v65; \cite{ott10}) employing the
standard unchopped long-range spectroscopy data processing script to
achieve spatially calibrated spectral maps.  We applied the
standard flux calibration scheme, where the system response applicable
to an observation ID is estimated from a calibration block at key
wavelengths. The calibration block is executed at the beginning of
each on- and off-source observation and the derived system response is
adopted for the duration of the observation. The absolute flux level
is propagated from the key wavelengths to the covered wavelengths
range applying the relative spectral response function (RSRF).  Unchopped raw
data suffer from instantaneous system response drifts due to
cosmic-ray impacts on detectors. Unlike in the case of chop/nod
observing mode where drifts are largely eliminated at the chopping
frequency in the chop\_on-chop\_off differential signal, any drifts
occurring at timescales shorter than the observation itself remains a
property of raw unchopped data. As part of the standard pipeline
spectral flat-field task, the continuum sampled by each of the 16
spectral pixels was scaled to the median value to correct for residual
flat field effects and to take advantage of redundancy in the data to
minimise the impact of eventual drifting pixels. In the final step,
flux calibrated cubes are created by subtracting on-off pairs to
completely remove the sky and telescope background. On each individual
off-subtracted spectral cubes we apply a multi-resolution wavelet
based continuum subtraction before data are projected into the final
hypersectral cubes. Using this adaptive technique any post-calibration
residual continuum wiggles broader than ~5-7$\times$ the width of an
unresolved line can be efficiently eliminated from the spectra.

\begin{figure}
\centering
\includegraphics[scale=0.45]{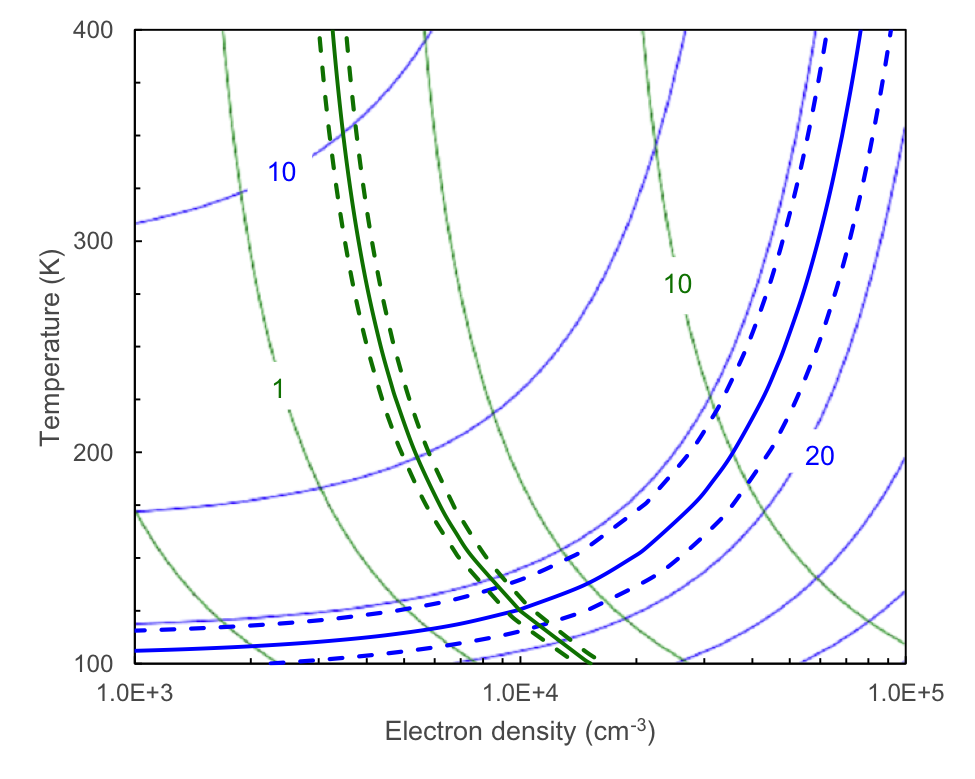}\\
\includegraphics[scale=0.45]{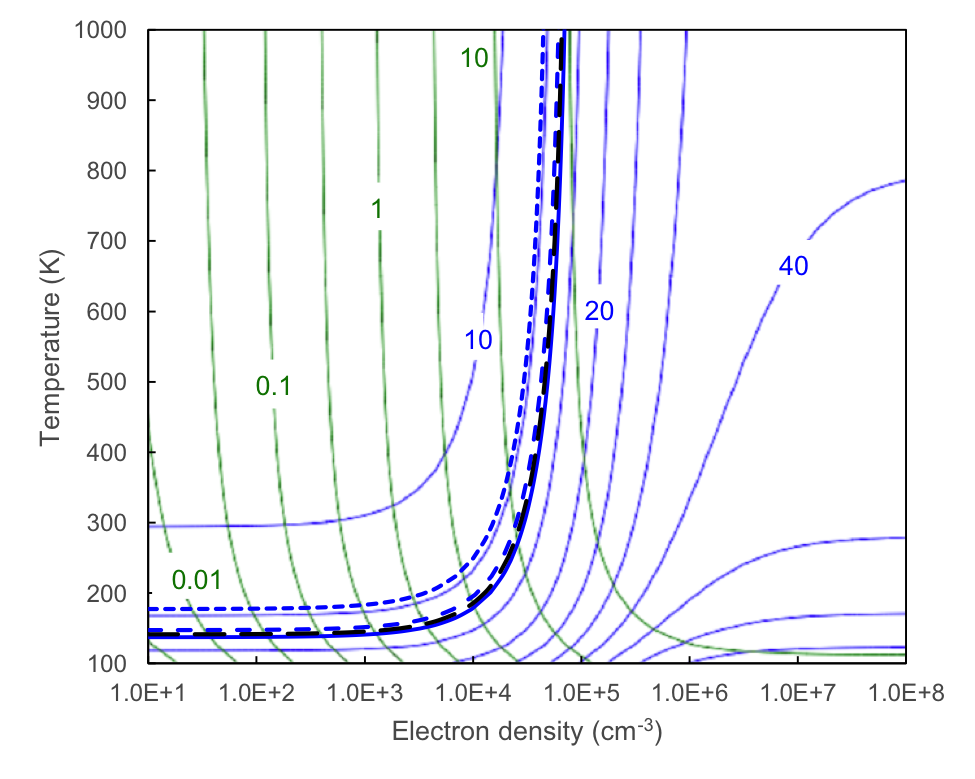}
\caption{{\it (Top)}: observed [O~I] 63.2/145.5 (blue) and [O~I] 63.2/[C II] 157.7  line ratio (green) contours compared to expectations for a single-component model. The observed ratios for the regions immediately outside the FIR~3 jet are indicated with heavy solid contours, and the $\pm$1 $\sigma$ band by dashed contours. Thin contours indicate the constant line ratios, covering the plotted range of density and temperature, with logarithmically-spaced contours.  The families of curves intersect near $n_e$~=~10$^{4}$ \percc~and T = 130 K, which lie in the ranges common for photodissociation regions in the Orion molecular-cloud complex.  {\it (Bottom)}:  [O~I] 63.2/145.5 intensity ratios observed along the jet plotted as heavy blue curves: FIR~4 (shortest dashes), FIR~3 (solid), SW/NE of FIR~3 (medium/long dashes). At high temperatures the intensity ratios would indicate electron density in the range 5$\times$10$^4$~-~8$\times$10$^4$ \percc.}
\label{figb1}
\end{figure}

\subsection{CO observations}

We also obtained CO J=6-5 map of the OMC-2 region made with the CHAMP+ seven element receiver array on the APEX~12 meter telescope.  The CO map has 9\arcsec~resolution similar to the 63~$\mu$m [O~I] map.  The details of these observations will be published in a forthcoming paper on a CHAMP survey of several Orion protostars.

%%%
\section{Analysis and results} \label{sec:ar}
%\subsection{Extended emission from PDRs}
The [O~I] 63~\micron~line map of the OMC-2 region is presented in Figure~\ref{fig1}. 
Inspection of the line map reveals a jet:  a narrow spatially collimated emission feature connecting FIR~3 and FIR~4. The jet feature is seen superposed on an uniform extended emission. In the following we show that the extended background component is consistent with the emission from a photodissociation region (PDR) on the surface of the OMC-2 cloud.

\subsection{Extended PDR emission in OMC-2}  \label{sec:ne}
The spatial morphology of the fine structure line emission indicates the presence of uniform extended emission in the OMC-2 region. Figure~\ref{fig_oc} shows the [O~I] 63~\micron~to [CII]  and  [O~I] 63 to 145 \micron~line ratio maps. The line ratios of the extended emission are approximately constant and distinct from the narrow jet-like emission, indicating a different physical origin for the extended emission. The OMC-2 region lies between the Orion Nebula Cluster (ONC) to the south and NGC~1977 to the north-east, both of which contain OB stars \citep{peterson08} and the UV radiation from which can produce photodissociation regions (PDRs) on the surface of molecular clouds. 
%show that the line ratios for the extended background emission are more or less constant and very different from that for the narrow jet-like emission, indicating the presence of two different physical components. 
%The narrow ridge like emission feature seen in [O~I] spatially coincides with the extended molecular outflow driven by FIR~3  \citep{williams03, shimajiri08} and most likely traces hot gas shocked by the jets responsible for the outflow.
%In order to confirm the presence of the extended PDR component, we analysed the line ratio maps of [O~I] 63 and 145 \micron~lines and [O~I] 63 \micron~and [CII] 158 \micron~lines, which are shown in Figure~\ref{fig2}.  

To quantify the emission from the extended component, we measured the line ratios in six 15\arcsec $\times$ 15\arcsec regions, three on either side of the jet.  The line ratios are remarkably constant between the boxes: \\
	${[\mathrm{O~I}] 63}/{[\mathrm{C~II}] 158} = 1.9 \pm 0.1$ , \\
	${[\mathrm{O~I}] 63}/{[\mathrm{O~I}] 146} = 17 \pm 1$.\\
%\begin{displaymath}   
%\centering
%	\frac{[\mathrm{O I}]63}{[\mathrm{C II}]158} = 1.9 \pm 0.1 \; \;  \mathrm{and} \;\; \frac{[\mathrm{O I}]63}{[\mathrm{O I}]146} = 17 \pm 1
% \end{displaymath}   
We compared these line ratios to those expected for a cloud with uniform density and temperature in order to estimate the physical conditions of the extended emission region.
If we imagine each [O~I] and [C II] line-emission region to have a uniform density and temperature, we can derive useful characteristic densities and temperatures from the line intensity ratios, under the common assumptions that the lines are optically thin and unextinguished, the atomic energy levels are collisionally excited, and the energy-level populations are in steady state \citep[e.g.,][]{of06}. 

Figure~\ref{figb1} contains results of such calculations for the intensity ratios we observe for the FIR~3 jet and for the adjacent cloud emission. We considered only the five lowest-energy levels for neutral oxygen, and the two lowest-energy levels for ionized carbon, to have significant population. We invoked only electrons as collision partners, as these rates dominate those from protons and hydrogen even when the ionization fraction is small. We took the relative abundance of C$^{+}$ and O to be the solar-system C/O ratio, 0.46 \citep{lodders09}. We used the collision strengths computed by \citet{bell98} and \citet{zat03} for $e-$O, and by \citet{tayal08} for $e-$C$^{+}$. Finally, we used the Einstein A-coefficients computed by Froese Fischer \& Tachiev (2004) and \citet{wiese07} for [O~I] and [C II] respectively.

The observed line ratios give an electron density of $n_e$$\sim$10$^{4}$~\percc ~and T = 130 K for the extended emission component (see Figure~\ref{figb1}), appropriate for PDRs with incident FUV field strength of $G_0 \sim$~100 \citep{kauf99}. The values for $n_e$, T and $G_0$ that we obtain lie in the ranges commonly observed for PDRs on the surface of molecular clouds \citep{ht97,ht99}, indicating that the extended component surrounding the jet is consistent with the emission from a photodissociation region (PDR) on the surface of the OMC-2 cloud.

\begin{figure}
   \centering
\includegraphics[scale=0.35]{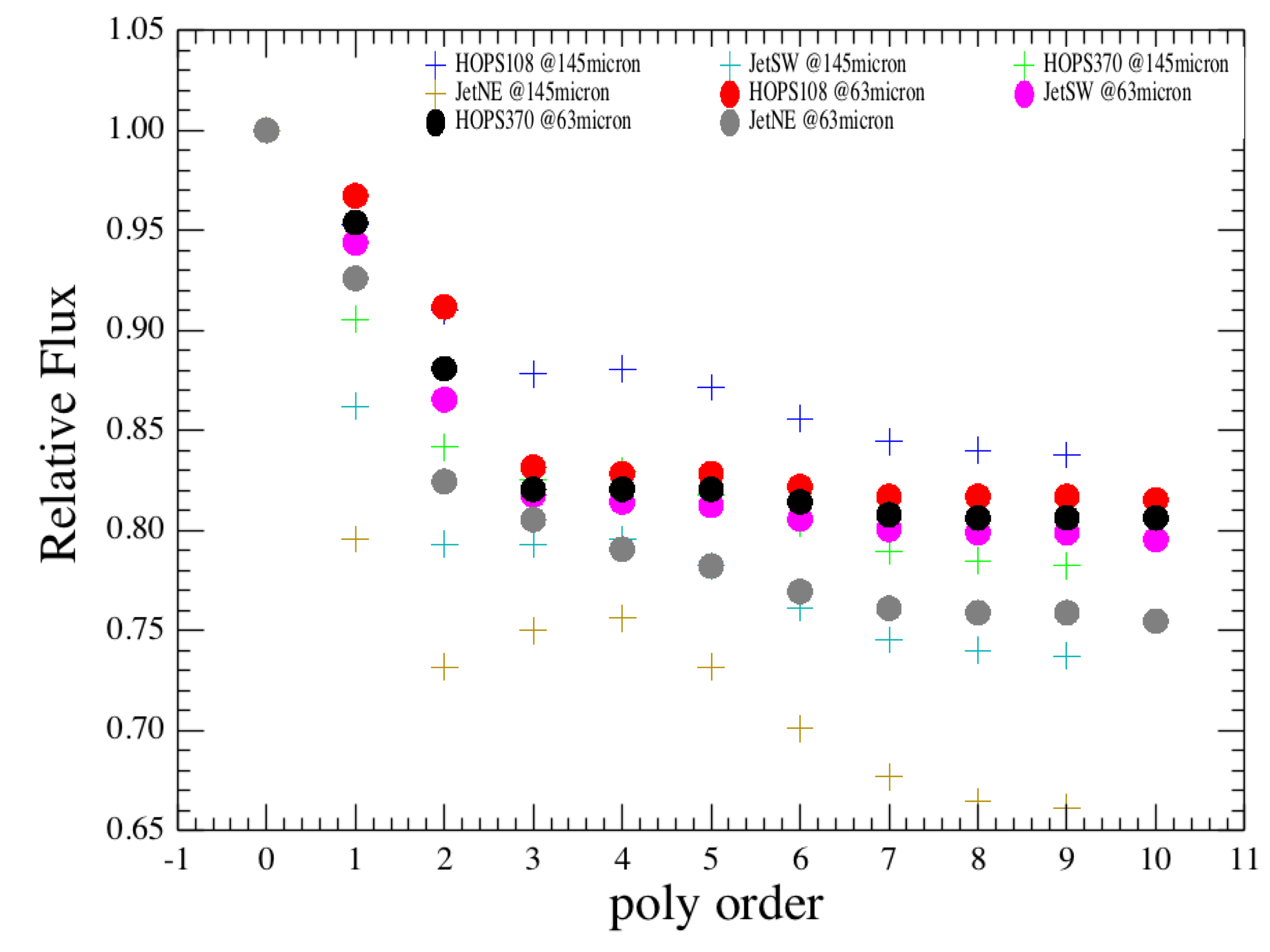}
\caption{ Flux from apertures along the jet for different polynomial order fits relative to zero order fit shown as a function of the polynomial order used for subtracting the extended background emission component.\label{fig_poly} }
%{\color{red} { [Do we need to show all the positions along the jet?  Co-authors, please comment]}}\label{fig_poly} }
\end{figure}

%%The \o1c2 and
%\o1o1 line ratio maps (see Figure~\ref{fig2} Appendix~\ref{sec:ne}) show that the line ratios of the extended emission are approximately
%constant and distinct from the narrow jet-like emission, indicating the presence of two different physical origins for the line emission:
%an extended photodissociation region (PDR) component and a collimated jet. 
\subsection{Separation of the extended PDR and the jet components}

We isolated the jet component from the PDR background for further analysis. We first extracted the extended emission component by fitting the line image with a third order polynomial after masking the central ridge like emission component above 30\% of the peak emission detected in the mapped area.
We used a third order polynomial to fit the background because it was found to be optimal. Higher order polynomial fits did not improve the background subtraction significantly. This is illustrated in Figure~\ref{fig_poly}, where we show the flux extracted from different apertures on the jet after subtracting the background by fitting polynomials of orders 0 to 10 divided by the flux extracted at the 0 order polynomial. The flux obtained for different polynomial orders are shown relative to that for the zeroth order polynomial fit.  For both the [O~I] 63~\micron~and [O~I] 145~\micron~ there is no significant improvement beyond a 3$^{rd}$ polynomial fit to the background emission.

\begin{figure*}
\begin{tabular}{cc}
   \centering
\includegraphics[scale=0.3]{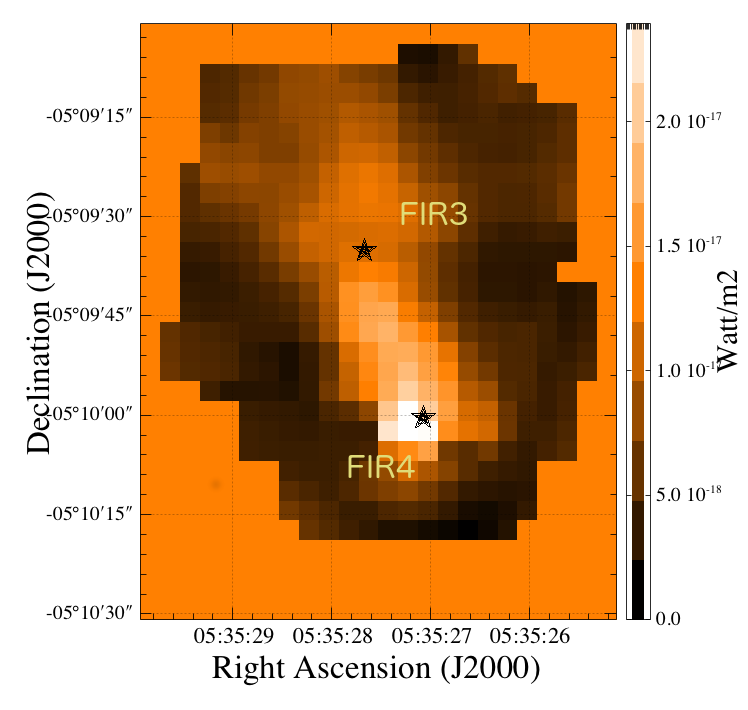} &
\includegraphics[scale=0.3]{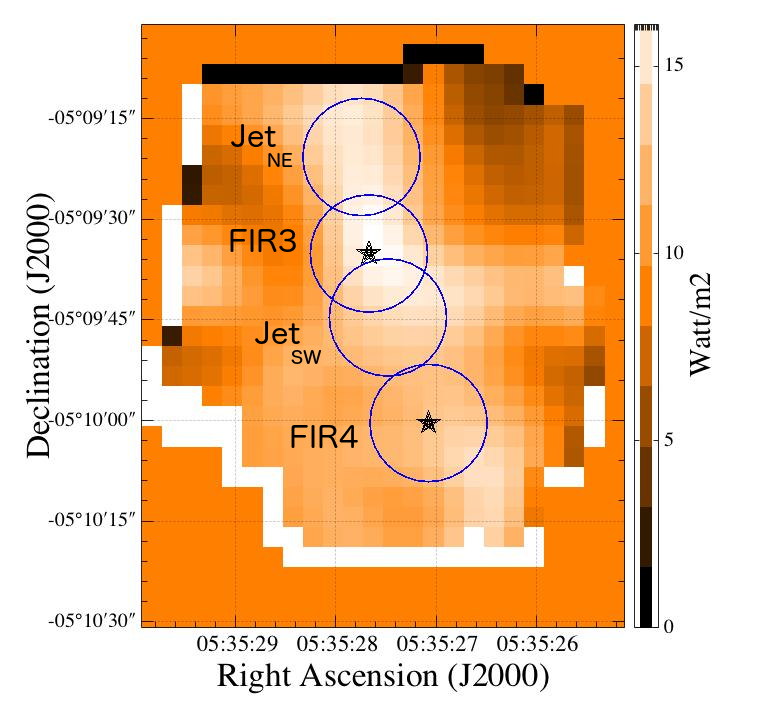} \\
\end{tabular}
\caption{ {\it (Left)}: [O~I] 145~\micron~line map  and {\it (Right)}: [O~I] 63 /  [O~I] 145 line ratio map, both after the subtraction of the extended component. \label{fig_o145} }
\end{figure*}

 We then subtracted the extended emission component from the total emission map.  Figure~\ref{fig1} shows the total [O~I] 63~\micron~line emission map, the background component and the residual jet component after subtracting the extended background.  The procedure was repeated for [O~I]~145~\micron~and the results are shown in Figure~\ref{fig_o145}. Flux values on the jet component have been summed up within 9\arcsec synthetic radii apertures. The central locations of the four apertures are indicated on Figure~\ref{fig1} and the line fluxes are listed in Table~\ref{tab2}. The statistical uncertainties in the extracted line fluxes were estimated from the fluctuations in the background-subtracted line images measured at several positions located off the jet.  These include {\it rms} fluctuations due to both the noise in the observed line and the residual from the PDR subtraction. The {\it rms}  noise of the [O~I] 63~\micron~line map is found to be 2.6~$\times$~10$^{-17}$~Wm$^{-2}$ per pixel while that for [O~I] 145~\micron~line map is 3.5~$\times$~10$^{-18}$~Wm$^{-2}$ per pixel. Uncertainties corresponding to the line flux integrated over 9\arcsec~radius apertures are listed in Table~\ref{tab2}.

%In order to calculate line
%sensitivities, we have masked the spectral lines $\pm$ 2 $\times$ the
%size of a resolution element on both sides of the line with respect
%the line centre. Taking advantage of the broad-range coverage from the
%spectral scans and the flat baseline subtracted continuum, we can
%derive a robust continuum noise estimate from the remaining unmasked
%~100 spectral bins around the line of interest. The bin-to-bin rms
%continuum noise estimate does not require further correction since our
%choice of spectral binning (upsample=1) ensures that the noise should
%not be strongly correlated between adjacent bins. The 3-sigma line
%sensitivities are calculated from the integrated flux of an unresolved
%emission line centered at the observed line wavelength and its line
%peak intensity is adjusted to the 3-sigma continuum noise estimate.
%{\color{blue}{[ Bea: Add a brief discussion on the method used to update uncertainties in the line flux: The uncertainty in the flux could be separated into intrinsic errors to the measurement plus calibration error. The first one have been calculated at the square root of three terms sum in quadratic: (1) the  RMS measured at the RMS maps, at the 4 target positions, the standard deviation of the bg measured at 5 positions outside the target-jet system and the standard deviation at the 9 positions (4 target and 5 sky positions) of the bg subtracted image. For the luminosity, mass-loss rate, ratios and electron density uncertainties, standard error propagation was done]}}

\subsection{The jet morphology}

\begin{figure*}
\centering
    \includegraphics[scale=0.6,trim=20 0 0 0]{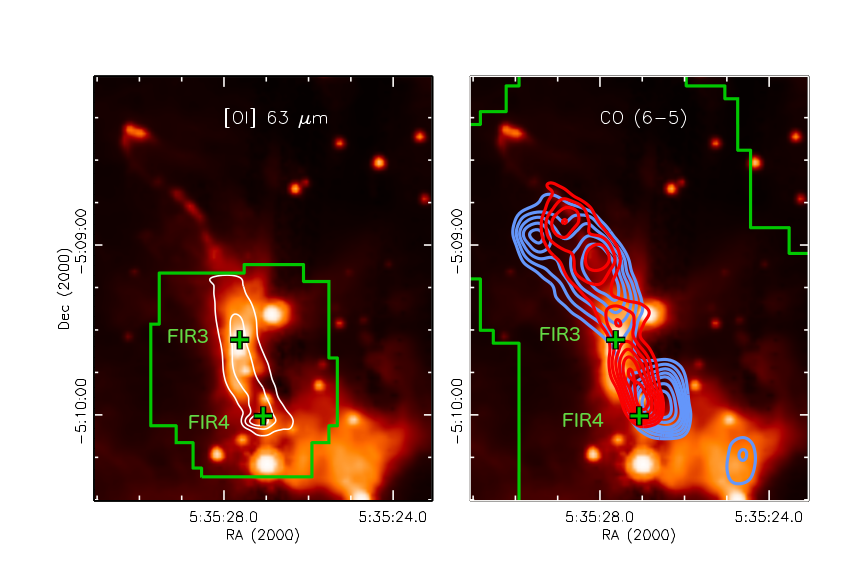}
   \caption{ {\it (Left)}: [O~I] 63 $\mu$m line emission (white contours; levels = 1, 2 \& 3 $\times$10$^{-16}$ Wm$^{-2}$/pixel) with the FOV outlined in green. {\it (Right)}: blue ($-$10 to 9~\kms) and red (14 to 30~\kms) wings of the CO 6-5 are the blue and red contours.  Both are overlaid on the 4.5 $\mu$m image, which shows the jet extending to the north-east.}  
      \label{fig4}
\end{figure*}

The  [O~I]  jet has a north-east to south-west orientation.  The jet emission extends from the edge of the mapped area in the north-east, through
FIR~3, and finally terminates at the location of FIR~4 in the south-west, where the [O~I] emission is the brightest.  Figure~\ref{fig4} shows contours of the [O~I] 63 \micron~jet on a {\it Spitzer}/IRAC 4.5~\micron~image \citep{megeath12}. The 4.5~\micron~band contains bright H$_{\mathrm 2}$ emission which traces the shocked molecular gas excited by the jet. We observe a prominent extended 4.5~\micron~jet terminating in a bow shock to the north-east of FIR-3 (extending past the PACS line map coverage). The 4.5~\micron~jet emission clearly appears to be driven by FIR~3; however, the counter-jet traced by [O~I] is not readily seen at 4.5~\micron. Figure~\ref{fig4} also shows an overlay of blue- and red-shifted CO
(6-5) outflow emission which traces the molecular gas entrained by the jet. The CO emission clearly follows the north-east component of the
4.5~\micron~jet as well as the [O~I] emission extending to the south-west and terminating at FIR~4. Since both CO outflow lobes have high-velocity line wings on both sides of the systemic velocity, the outflow must lie close to the plane of the sky. The combined morphology of all
three emission components ([O~I], 4.5~\micron, and CO)  strongly suggests that the jet/outflow is driven by FIR~3.

\subsection{Excitation conditions in the jet}
To investigate the physical conditions along the jet, we measured the [O~I] 63 \micron/145 \micron~line ratio at four different locations along the jet (Table~\ref{tab2}).  We compare the observed ratios to predictions for a cloud with uniform density and temperature, described in  Section~\ref{sec:ne} and Figure~\ref{figb1}. At the high temperatures expected in the dissociative shocks in jets, the line ratio is insensitive to the precise temperature, and indicates electron density ($n_e$) in the range of 5-8$\times$10$^{4}$ \percc~(see Figure~\ref{figb1}).  The $n_e$ values for the four different sections of the jet were computed for T = 3000 K (see Table~\ref{tab2}).  The densities that we derive appear to suggest a gradient along the jet with the density of the jet decreasing with increasing distance from FIR~3. However, the variations in the density along the jet are not significantly above the associated uncertainties, so the presence of a density gradient in the jet cannot be established.

%The $n_e$ values decreases towards the south-west, reaching a minimum value at the [O~I] line emission peak close to FIR~4.  The density also drops on the opposite side, to the north-east of FIR-3.  The $n_e$ values suggest that the [O~I] jet is driven by FIR~3 with the maximum density of the jet at the location of the driving source, as expected if the [O~I] emission originates in the decelerated jet material.

 \begin{table*}
 {\small
%\centering
\caption{Observed properties of the [O~I] jet from FIR~3 \label{tab2} } 
\begin{tabular}{c c c c c c c c} \hline  
Position & RA(J2000)&Dec(J2000) & \multicolumn{2}{c}{  [O~I] 63~\micron~~\tablefootmark{a}}  & Mass-loss rate  & [O~I] 63~\micron/145~\micron  & Electron density \\ 
  along the jet            & & & Line flux   & Line luminosity & $\dot{M}_{out}$ & flux ratio~~\tablefootmark{b}  & $n_{e}$  \\
             & & & ($\times$ 10$^{-15}$ Wm$^{-2}$)& ($\times$ 10$^{-2}$ L$_{\odot}$) &  ($\times$ 10$^{-6}$ M$_{\odot}$yr$^{-1}$) &  & ($\times$ 10$^4$~\percc)   \\ \hline
FIR~4                  & 5~35~27.07 & $-$5~10~0.37 & 5.1~$\pm$~0.14     & 2.8              &  2.3  & 12.2~$\pm$~0.7 & 4.8~$\pm$~0.7 \\
Jet$_{\mathrm{SW}}$& 5~35~27.48 & $-$5~09~44.64 & 5.2~$\pm$~0.14   & 2.8              &   2.3 & 13.5~$\pm$~0.8 & 6.5~$\pm$~0.9 \\
FIR~3                 & 5~35~27.67 & $-$5~09~35.10 & 4.8~$\pm$~0.14  & 2.6            &  2.1  & 14.1~$\pm$~1.0 & 7.4~$\pm$~1.1\\
Jet$_{\mathrm{NE}}$ & 5~35~27.74 & $-$5~09~20.70 & 3.8~$\pm$~0.14   & 2.1                & 1.7 & 13.8~$\pm$~1.2 & 7.0~$\pm$~1.4 \\
\tablefootmark{c}Entire jet                     &                     &                           & 17.3$\pm$0.3    & 9.5                &  7.7  &     &     \\
 \hline
\end{tabular}
\tablefoot{  \\
\tablefoottext{a}{The [O~I]~63 line fluxes are extracted using a 9\arcsec~radius aperture at the indicated positions. The statistical uncertainties in the line flux are estimated from the {\it rms} noise in the background-subtracted line maps. The line luminosities and mass loss rates have the same fractional uncertainty as the line flux fractional uncertainty.  The systematic uncertainties in the line flux for unchopped observations are dominated by the calibration errors and are expected to be $\la$10-12\%.}\\
\tablefoottext{b}{The line ratio maps are calculated at matched angular resolution and the values presented here are extracted using a 9\arcsec~radius aperture at the indicated positions. The flux ratio uncertainties are calculated from the statistical uncertainties in the fluxes. }\\
\tablefoottext{c}{This corresponds to the total flux from the [O~I]~63~\micron~line map (after the subtraction of the extended emission component) obtained by adding up all the pixels with flux values~$\ga$~3$-\sigma$, where $\sigma$ is the {\it rms} of the map.
%{\color{red} { [The corresponding mass loss rate and the implied accretion rate (for $b$ = 0.1) are too high. This would imply a \lacc~of  1800 \lsun, with the assumptions made in Section 3.5. This is much higher than the \lbol~of even FIR~3.  Co-authors, please comment on this. Should we keep this at all ?]}}
}
}
}
\end{table*}

\subsection{Mass-loss rates from jets}
The [O~I] 63 \micron~line luminosity provides a direct measure of the mass loss rates from protostars \citep{hollen85, hm89}. This line is the primary coolant of the predominantly atomic gas in the postshock gas for temperatures in the range of 5000-500 K. The [O~I] line luminosity of a planar shocked region is proportional to the rate at which jet material flows into the shock front, or the mass loss rate of the driving source, given by \\
$\dot{M}_{out}=8.1\times10^{-5} (L_{[O~I]}/L_{\odot})~\textrm{\msunyr}$. \\
%\[ \dot{M}_{out}~=~10^{-4}\;\frac{L_{[O~I]}}{L_{\odot}}~\textrm{\msunyr}\]
Presuming a protostar's mass loss rate to be constant over the dynamical time of the jet, and neglecting emission from shocks within the jet, the total [O~I] 63~\micron~luminosity,  seen from a protostar's outflow thus yields the protostar's mass-loss rate. With these caveats, the mass loss rate for FIR3, corresponding to the total [O~I]~63~\micron~luminosity we observe in the jet is 7.7$\times$10$^{-6}$~\msunyr. The total [O~I] flux was measured from the PDR subtracted line map by adding up all the pixels with flux values~$\ge$~3~$\times$~the {\it rms} of the map (Table~\ref{tab2}). We also computed the shock flow rates at four different locations along the jet from the [O~I] luminosities measured within 9\arcsec of the positions listed in Table~\ref{tab2} (also see Figure~\ref{fig1}). The flow rates at different sections of the jets lie within a narrow range of 1.7-2.3 $\times$10$^{-6}$~\msunyr~and differ only by 26\% at most, indicating that  the mass flow along the jet is relatively smooth and continuous.

Mass-loss rates from protostars are closely connected to mass accretion rate onto protostars: theoretical models of jet launching predict a proportionality between mass loss rate and mass accretion rate,\\ $\dot{M}_{out}$~=~$b\:\dot{M}_{acc}$,\\ where $b$ has a theoretical as well as a mean observed value near 0.1 \citep[][]{pp92,watson15}.  The mass flow rate of $\dot{M}_{out}$~=~2.3$\times$10$^{-6}$~\msunyr~at the location of FIR~4 implies a mass accretion rate, \macc~=~2.3$\times$10$^{-5}$~\msunyr, which corresponds to an accretion-luminosity, \lacc~=~350~\lsun, assuming \\ $R_{\star}$/$M_{\star}$~=~1~$R_{\odot}$/$M_{\odot}$,\\ appropriate for intermediate-mass young stars on the birth line \citep{pallastahler93}. This value of \lacc~is close to a factor of 10 higher than the observed \lbol~for FIR~4. Further, as noted by \citet{furlan14}, the true luminosity of FIR~4 could be as small as 14 \lsun. On the other hand, the \lacc~that we derive is consistent with the observed \lbol~for FIR~3.  Moreover, the \macc~values indicated by the mass loss rates derived from [O~I] is remarkably close to the envelope infall rate of 1$\times$10$^{-5}$~\msunyr~estimated for FIR~3 from the radiative transfer modeling of the observed broad band spectral energy distribution \citep{adams12, furlan16}. All these different lines of argument strongly suggest that the [O~I] jet is driven by FIR~3.

\section{Discussion}
Our analysis indicates that the [O~I] jet connecting FIR~3 and FIR~4 actually originates from the FIR~3 protostar.  Although our \herschel/PACS line maps do not cover the full extent of the north-eastern portion of the 4.5~\micron~shocked emission, the [O~I] jet is observed on either side of FIR~3 and terminates at FIR~4.  The jet is also spatially aligned with the CO~(6-5) outflow, which extends to FIR~4 in the south. 
%The observed  [O~I]~63~\micron~/~145~\micron~line ratios at different parts of the extended jet show the density to be progressively decreasing along the jet on either side of FIR~3. Since the [O~I] emission predominantly arises from the shocked jet material, such a density gradient is expected if the jet is driven by FIR~3.  
The near constant mass loss rate found along the [O~I] jet, indicating a steady and continuous mass flow, also suggests that FIR~3 is driving the jet.  The mass flow rate close to FIR~3 and 20\arcsec (8400 AU) away at the location of FIR~4 are very similar. It is highly unlikely that FIR~4, which is $\sim$ 10 times less luminous than FIR~3 is driving a jet with mass-flow rate similar to that of FIR~3. These pieces of evidence strongly suggest that the driving source of the observed [O~I] jet is FIR~3.

The peak [O~I] emission, nevertheless,  is observed close to FIR~4.  All the high-excitation molecular cooling lines in the far-IR (H$_2$O, CO and OH) observed with \herschel/PACS also have their peak emission at the location of FIR~4 \citep[][Manoj et al. in prep]{furlan14}. In fact, FIR~4 is one of the brightest far-IR line emitting source among all the protostars observed by HOPS. The far-IR line luminosities measured within a 9.4\arcsec aperture around FIR~4 is more than 4$\times$ higher than that observed for FIR~3 \citep{manoj13}.  Our analysis and other studies of the region, however,  do not show any evidence for FIR~4 driving a powerful outflow/jet which could excite such intense line emission. 
We argue, therefore, that the bright line emission seen toward FIR~4 is the terminal shock (Mach disk) of the FIR~3 jet. Such terminal shocks are often the brightest features of YSO jets, and comprise many of the best-known Herbig-Haro objects.  FIR~4 may simply lie along the line of sight and may not even be physically associated with the shocked emitting region.

The jet driven by FIR~3 is the first to be imaged in [O~I]~63~\micron~from an intermediate-mass protostar.  [O~I] jets from low mass class~0 protostars observed with \herschel/PACS by \citet{nisini15} show  mass-loss rates in the range of 1-4~$\times$10$^{-7}$~\msunyr.  These mass-loss rates are more than an order of magnitude lower than that of FIR~3.  The mass loss rate for FIR~3 is higher than even the relatively high mass-loss rate of 2-4~$\times$10$^{-6}$~\msunyr~found for HH~46 by \citet{nisini15}.  Thus our results indicate that the mass-loss rates in intermediate-mass protostars are $\ga$ 10 times higher than that observed for most low-mass class 0 protostars, consistent with the idea that they drive more powerful jets than their low-mass counterparts.

\begin{acknowledgements}
      Part of this work was supported by ISDEFE. Thanks to Katrina Exter, Jeroen de Jong and Pablo Riviere-Marichalar for their support at PACS data processing. Support was provided by National Aeronautics and Space Administration (NASA) through awards issued by the Jet Propulsion Laboratory, California Institute of Technology (JPL/Caltech). M.O. and A.K.D.R. acknowledge support from MINECO (Spain) AYA2011-3O228-CO3-01 and AYA2014-57369-C3-3-P grants (co-funded with FEDER
funds). We include data from \herschel, a European Space Agency space observatory with science instruments provided by European-led consortia and with important participation from NASA. We also use data from the Spitzer Space Telescope, operated by JPL/Caltech under a contract with NASA, and APEX, a collaboration between the Max-Planck-Institut f\"ur Radioastronomie, the European Southern Observatory, and the Onsala Space Observatory.
\end{acknowledgements}

\end{document}